\documentstyle[12pt]{article}
  \let\g=\gamma \let\d=\delta
\let\e=\varepsilon  \let\h=\eta 
  \let\l=\lambda \let\m=\mu
\let\n=\nu \let\x=\xi \let\p=\pi

  \let\P=\Pi 
  \let\D=\Delta 
\def\mn{{\m\n}}

\def\0{\over } \def\1{\vec } \def\2{{1\over2}} \def\4{{1\over4}}
\def\5{\bar } %\def\5{\overline }
\def\6{\partial }
\def\7#1{{#1}\llap{/}}
\def\8#1{{\textstyle{#1}}} \def\9#1{{\bf {#1}}}

\def\({\left(} \def\){\right)} \def\<{\langle } \def\>{\rangle }
\def\[{\left[} \def\]{\right]}  

\def\AP#1#2#3{Ann. Phys. (NY) {\bf #1}, #2 (19#3)}
\def\ZETF#1#2#3{Zh. Eksp. Teor. Fiz. {\bf #1}, #2 (19#3)}
\def\JETP#1#2#3{Sov. Phys. JETP {\bf #1}, #2 (19#3)}
\def\YF#1#2#3{Yad. Fiz. {\bf #1}, #2 (19#3)}
\def\SJNP#1#2#3{Sov. J. Nucl. Phys. {\bf #1}, #2 (19#3)}

\def\PLB#1#2#3{Phys. Lett. {\bf B#1}, #2 (19#3)}
\def\PRL#1#2#3{Phys. Rev. Lett. {\bf #1}, #2 (19#3)}
\def\RMP#1#2#3{Rev. Mod. Phys. {\bf#1}, #2 (19#3)}

\def\PRD#1#2#3{Phys. Rev. D {\bf #1}, #2 (19#3)}

\def\NPB#1#2#3{Nucl. Phys. {\bf B#1}, #2 (19#3)}

\def\be{\begin{equation}}
\def\ee{\end{equation}}
\begin{document}
\begin{titlepage}
\begin{flushright}
CERN-TH.6375/92\\
\end{flushright}
\vfill
\begin{center}
{\Large \bf Comment on
``Damping of energetic gluons and
quarks in high-temperature QCD''}\\
\vfill
{\large Anton Rebhan}\\
\bigskip
{\sl
Theory Division, CERN\\
CH-1211 Geneva 23, Switzerland\\}
\vfill
{\large ABSTRACT}
\end{center}
\begin{quotation}
Burgess and Marini have recently pointed out that the leading
contribution to the damping rate of energetic gluons and
quarks in the QCD plasma, given by $\g=c g^2\ln(1/g)T$,
can be obtained by simple arguments
obviating the need of a fully resummed perturbation theory
as developed by Braaten and Pisarski.
Their calculation confirmed
previous results of Braaten and Pisarski,
but contradicted those proposed by Lebedev and Smilga.
While agreeing with the general considerations made by
Burgess and Marini, I correct their actual calculation of
the damping rates, which is based on a wrong expression for
the static limit of the resummed gluon propagator. The effect
of this, however, turns out to be
cancelled fortuitously
by another mistake,
so as to leave all of their conclusions unchanged.
I also verify the gauge independence of the results, which in the
corrected calculation arises in a less obvious manner.
\end{quotation}
\vfill
\begin{flushleft}
CERN-TH.6375/92\\
January 1992\\
\end{flushleft}
\end{titlepage}
\def\quref#1{[\ref{#1}]}
 \def\queq#1{(\ref{#1})}

It has been established by Braaten and Pisarski \quref{rBP}
that a perturbation theory for the dispersion relations of
quasi-particles in high-temperature QCD requires at least
resummation of the leading-order terms, called ``hard
thermal loops'', whose characteristic
scale is given by $gT$, where $g$ is the coupling constant and
$T$ the temperature. By now, a number of applications exist
[\ref{rP}--\ref{rBT}] which employ
the resummation techniques developed in Ref.~\quref{rBP}
to explore the physics of the hot QCD plasma at the scale $g^2T$.
Complete results can be obtained, if they are not sensitive
to a further resummation of the corrections of order $g^2 T$, which
would have to
include the perturbatively incalculable screening of static
magnetic fields \quref{rmm}.

Burgess and Marini \quref{rBM} have recently discussed the
case where resummation of the hard thermal loops leaves logarithmic
infrared divergences, and they have made precise the notion
\quref{rP} that the resummation procedure still allows to reliably
extract terms $\propto g^2 T \ln(m_{el.}/m_{magn.}) \sim
g^2 \ln(1/g) T$,
if not those of order $g^2T$.

The particular example considered in Ref.~\quref{rBM} is the
evaluation of the leading contributions to the damping rate
of gluons or quarks with momenta $|\9p|\gg gT$. This kinematical
region leads to an enormous simplification of the resummation
program, because only the leading corrections to one internal
propagator carrying soft integration momentum need be resummed,
with no complications from the vertices. The similar case of
very massive quarks has previously been
discussed in Refs.~\quref{rP}
and \quref{rBT}. Burgess and Marini further noticed that in such
processes, which are dominated by the subleading scale $g^2T$,
only the static limit of the resummed gauge propagator is needed.

The calculation thus becomes technically similar to the well-known
resummation of ``ring diagrams'' in thermodynamical potentials,
which goes under the name of ``plasmon effect''
\quref{rK}. However, this term is somewhat misleading, as only
the static limit of internal lines with multiple self-energy
%($\P_\mn}$)
insertions is relevant, which thus resums the electric Debye
screening mass rather than
the (different) plasmon mass corresponding to
long-wavelength plasma oscillations.
The latter is determined by the long-wavelength limit of the
gluon self-energy
\be
\lim_{\9q\to0} \P_\mn(q_0,\9q) = m^2 (\h_\mn-\d_\m^0 \d_\n^0)
+O(gmq_0),
\label{elwl}\ee
with $m^2={1\09}(C_a+\2n_q)(gT)^2$,
whereas the static limit is
\be
\lim_{q_0\to0} \P_\mn(q_0,\9q) = m^2_{el.} \d_\m^0 \d_\n^0
+O(gm\9q),
\label{estat}\ee
with $m^2_{el.}=3m^2$ \quref{rHTL}.
Evidently, these limits do not commute.

In the calculation carried out in Ref.~\quref{rBM}, Eq.~\queq{elwl}
was used instead of Eq.~\queq{estat}
for the resummed gluon propagator at zero frequency, which
led the authors of Ref.~\quref{rBM} to using
\be
\D_\mn^{*\,\hbox{\tiny wrong}}\Big|_{q_0=0}=
-\[ {1\0q^2}
\d_\m^0 \d_\n^0
+{1\0q^2-m^2}\(\h_\mn-
\d_\m^0 \d_\n^0-{q_\m q_\n\0q^2}\)+
\x{q_\m q_\n\0(q^2-\x m^2)q^2}\]
\label{eDwrong}\ee
in place of the correct one
\be
\D_\mn^{*}\Big|_{q_0=0}=
-\[ {1\0q^2-m_{el.}^2}
\d_\m^0 \d_\n^0
+{1\0q^2}\(\h_\mn-
\d_\m^0 \d_\n^0-{q_\m q_\n\0q^2}\)+\x{q_\m q_\n\0(q^2)^2}\].
\label{eD}\ee
In the latter only the spatially
longitudinal mode is screened, leaving
both the spatially transverse mode and the (4-D longitudinal)
gauge mode massless.

Recalculation of the ``hard'' ($|\9q|>\l\gg g^2T$)
contributions to the
damping rate $\g$ of energetic ($|\9p|\gg gT$) transverse gluons
considered in Ref.~\quref{rBM} leads to
\begin{eqnarray}
\g^{\hbox{\tiny hard}} &=& g^2 C_a T {1\04\p^2}
{\rm Im} \int_{-1}^1 dz
\int_{\l}^\infty{dq\,q\0z+q/2|\9p|-i\e}
\[{1\0q^2}-{1\0q^2+m_{el.}^2}-(1-\x){z^2\0q^2}\] \nonumber\\
&&\;+O(g^2T\l^0),
\label{eg}\end{eqnarray}
where the terms in the large brackets
correspond to the contributions
of spatially transverse, spatially longitudinal, and gauge modes,
respectively. [In the case of quarks, it turns out that
the only change consists in replacing $C_a$ by $C_f$.]

On the other hand, with the wrong propagator of Eq.~\queq{eDwrong}
used in Ref.~\quref{rBM}, these terms would read
\be
\[{1\0q^2+m^2}-{1\0q^2}-(1-\x){z^2q^2\0(q^2+m^2)(q^2+\x m^2)}\].
\label{egwrong}\ee

The leading contribution to $\g$
can be extracted from the
logarithmic dependence of $\g^{\hbox{\tiny hard}}$
on the cutoff $\l\ll gT$, together with the assumption
that the inherent scale of the undetermined soft contribution
is given by $g^2 T$ (through the non-perturbative magnetic
mass or through dynamical screening at this scale).
The spatially transverse and spatially longitudinal contributions
in Eq.~\queq{eg} thus lead to
\be
\g\approx {g^2 C_a T\04\p}\(\ln{m_{el.}\0\l}+\ln{\l\0g^2T}\)
= {g^2 C_a T\04\p}\ln{1\0g}+O(g^2T),
\label{egln}\ee
with the transverse mode being responsible for the dominant
term proportional
to $\ln(g^2T)$, and therefore for the positive sign of $\g$.
The latter is a consequence of
the positivity of the transverse density in
a spectral representation of the
resummed gluon propagator \quref{rP}.

The wrong result of Eq.~\queq{egwrong}, on the other hand,
should have led to a result of equal magnitude,
but with a reversed sign, as the roles of spatially longitudinal
and transverse modes happen to be interchanged.
(The difference between $m$ and $m_{el.}=\sqrt{3}m$
only affects the terms of $O(g^2)$.)
The fact that in Ref.~\quref{rBM} also a
positive result was reported is due to the additional mistake
of a reversed sign of $i\e$ in their Eq.~(11) compared with
Eq.~\queq{eg} above. With the usual sign convention $\g=-{\rm Im}\,
E|_{pole}$, the correct analytical continuation is given by
$k_0\to k_0+i\e$.

A more conspicuous difference between the correct and the wrong
results, Eq.~\queq{eg} and Eq.~\queq{egwrong}, respectively,
concerns the contributions from the gauge modes.
With the wrong expression for the static gluon propagator,
Eq.~\queq{eDwrong}, the gauge modes
obviously would not contribute to
the infrared singular part, whereas in the corrected result,
Eq.~\queq{eg}, they seem to do so by superficial power counting.
However, performing the angular integration
\be
\int_{-1}^1 dz {z^2\0z+q/2|\9p|-i\e} = O\({q\0|\9p|}\)
\ee
reveals that they indeed do not contribute to the leading
logarithms in Eq.~\queq{egln}, as expected from general
arguments \quref{rKKR} for the gauge independence of
dispersion relations in finite-temperature QCD.

Thus, all the results on $\g$ presented in
Ref.~\quref{rBM}, its magnitude,
its sign, and its gauge independence,
remain, somewhat fortuitously, unchanged, and continue to confirm
the results by Braaten and Pisarski \quref{rBPu}, while
contradicting those proposed by Lebedev and Smilga \quref{rLS}.

{\it Acknowledgements:} I should like to thank Tanguy Altherr and
Rob Pisarski for useful discussions.
\vfill\eject

\noindent{\bf References}
\newcounter{nom}
\begin{list}{[\arabic{nom}]}{\usecounter{nom}}
\item
E. Braaten and R. D. Pisarski, \PRL{64}{1338}{90};
\NPB{337}{569}{90}.
%NPB{339}{310}{90}.
\label{rBP}
\item
R. D. Pisarski,\PRL{63}{1129}{89}.
\label{rP}
\item
E. Braaten and R. D. Pisarski, \PRD{42}{2156}{90}.
\label{rBPg}
\item
E. Braaten and T. C. Yuan, \PRL{66}{2183}{91}.
\label{rBY}
\item
M. Thoma and M. Gyulassy, \NPB{351}{491}{91};
E. Braaten and M. H. Thoma, \PRD{44}{1298}{91}.
\label{rBT}
%
%\item
%G. Baym, H. Monien, C. J. Pethick and D. G. Ravenhall,
%\PRL{64}{1867}{90}.
%\label{rBMPR}
%
\item
A. D. Linde, \PLB{96}{289}{80};
D. Gross, R. Pisarski and L. Yaffe, \RMP{53}{43}{81}.
\label{rmm}
\item
C. P. Burgess and A. L. Marini, \PRD{45}{R17}{92}.
\label{rBM}
\item
J. Kapusta, {\it Finite Temperature Field Theory}
(Cambridge Univ. Press, Cambridge, England, 1985).
\label{rK}
\item
V. P. Silin, \ZETF{38}{1577}{60} [\JETP{11}{1136}{60}];
O. K. Kalashnikov and V. V. Klimov, \YF{31}{1357}{80}
 [\SJNP{31}{699}{80}];
%V. V. Klimov, \ZETF{82}{336}{82} [\JETP{55}{199}{82}];
H. A. Weldon, \PRD{26}{1394}{82}.
\label{rHTL}
\item
R. Kobes, G. Kunstatter and A. Rebhan, \PRL{64}{2992}{90};
\NPB{355}{1}{91}.
\label{rKKR}
\item
E. Braaten and R. D. Pisarski, unpublished.
\label{rBPu}
\item
V. V. Lebedev and A. V. Smilga, \AP{202}{229}{90};
 \PLB{253}{231}{91}.
\label{rLS}
\end{list}
\end{document}